\documentclass[aps,prd,floatfix,notitlepage,superscriptaddress,preprintnumbers,nofootinbib,11pt]{revtex4-2}
\usepackage{graphicx} 
\usepackage{amssymb}
\usepackage{amsmath}
\usepackage{color}
\usepackage{multirow}
\usepackage{pifont}
\newcommand{\xmark}{\ding{55}}%
\usepackage{natbib,setspace}
\definecolor{darkblue}{rgb}{0,0,0.5}
\usepackage[colorlinks,linkcolor=darkblue,citecolor=darkblue,urlcolor=darkblue]{hyperref}

\newcommand\nn{\nonumber}

\begin{document}

\title{Neutrino phenomenology in the modular $S_3$ seesaw model}

\author{Mitesh Kumar Behera}
\email{miteshbehera1304@gmail.com}
\affiliation{Department of Physics, Faculty of Science, Chulalongkorn University, Bangkok 10330, Thailand}

\author{Pawin Ittisamai}
\email{pawin.i@chula.ac.th}
\affiliation{Department of Physics, Faculty of Science, Chulalongkorn University, Bangkok 10330, Thailand}

\author{Chakrit Pongkitivanichkul}
\email{chakpo@kku.ac.th}
\affiliation{Khon Kaen Particle Physics and Cosmology Theory Group (KKPaCT), Department of Physics, Faculty of Science, Khon Kaen University, 123 Mitraphap Rd, Khon Kaen 40002, Thailand}

\author{Patipan Uttayarat}
\email{patipan@g.swu.ac.th}
\affiliation{Department of Physics, Srinakharinwirot University, 114 Sukhumvit 23 Rd., Wattana, Bangkok
10110, Thailand}

\begin{abstract}
We have studied neutrino phenomenology in the supersymmetric type-I seesaw model endowed with the $\Gamma_2 \simeq S_3$ modular symmetry. We have identified different realizations of the $S_3$ modular symmetry, referred to as models A, B, C, and D. The 4 models are compatible with neutrino mass being inverted ordering (IO). Moreover, models A, B, and D can also accommodate normal ordering (NO) neutrino masses. We identify parameter space for each model compatible with neutrino oscillation at the 2-$\sigma$ level. We then proceed to study the neutrino phenomenology of each model. We find that the lightest neutrino mass can be as light as 0.64 meV in the case of NO in model A and 50 meV in the case of IO in model D. The smallest effective electron neutrino mass attainable in our analysis is 8.8 meV in the case of NO (model A), and 50 meV for IO (model D). Finally, we note that the effective Majorana mass can be as small as 0.33 meV in the case of NO (model A) and 22 meV for IO (model D).
\end{abstract}

\maketitle

\section{Introduction}
The Standard Model (SM) initially posited neutrinos as massless particles, yet experimental evidence from neutrino oscillation data suggests otherwise, indicating they possess tiny but non-zero masses. This revelation solidifies the presence of neutrino mixing, implying that at least two neutrinos have non-zero masses~\cite{King:2003jb, Altarelli:2004za, Mohapatra:2006gs}. Unlike other fermions in the SM, neutrinos lack right-handed counterparts, precluding them from acquiring masses through the Higgs mechanism. However, the introduction of a dimension-five Weinberg operator~\cite{Weinberg:1979sa, Wilczek:1979hc,Weinberg:1980bf} offers a plausible mechanism for neutrino mass generation, although the origin and flavor structure of this operator remains debated. Hence, exploring scenarios beyond the SM (BSM) becomes essential to accommodate the observed non-zero neutrino masses.

Various models have been proposed to elucidate neutrino masses and mixing patterns, including the seesaw model~\cite{Minkowski:1977sc, Mohapatra:1979ia, Gell-Mann:1979vob,Yanagida:1979as,Glashow:1979nm,Chen:2009um}, radiative neutrino mass generation~\cite{Zee:1980ai,Zee:1985id,Babu:1988ki,Babu:2013pma,Nomura:2019dhw,Primulando:2022lpj}, and models involving extra dimensions~\cite{Arkani-Hamed:1998wuz, Carena:2017qhd, Forero:2022skg, Roy:2023dyq, Anchordoqui:2023wkm}. Understanding neutrino mixing patterns has prompted the adoption of symmetry-based approaches, wherein discrete flavor symmetries~\cite{Altarelli:2005yp,Altarelli:2010gt,Ishimori:2010au,Hernandez:2012ra,King:2013eh,King:2014nza,Hagedorn:2017zks,Patel:2023voj}, are introduced to enforce specific mixing patterns among leptons and possibly other fields. While these approaches have some early successes, recent experimental data have revealed their deficiencies, requiring departures from minimal models. Incorporating the measured value of parameters such as the reactor angle ($\theta_{13}$) and CP-violating phase often leads to non-trivial corrections and challenging predictability. Therefore, new methods are to be explored to tackle the above drawbacks, to which modular symmetry serves the purpose.

Modular flavor symmetry~\cite{Leontaris:1997vw,Kobayashi:2018vbk,feruglio2019neutrino,deAdelhartToorop:2011re} offers a new tool for model building. It promotes Yukawa couplings and mass parameters to modular forms, which transform in a non-trivial representation of a flavor symmetry group. The idea of modular flavor symmetry have been widely applied to construct viable neutrino mass models with symmetry groups such as  $S_3$~\cite{Kobayashi:2018wkl,Okada:2019xqk,Meloni:2023aru,Mishra:2020gxg}, $A_4$~\cite{Nomura:2023usj,RickyDevi:2024ijc,Kumar:2023moh,Mishra:2022egy,Wang:2019xbo,Kashav:2021zir,Kashav:2022kpk,Mishra:2023ekx,Lu:2019vgm, Kobayashi:2019gtp,Nomura:2019xsb,Behera:2020lpd,MiteshKumar:2023hpg,Nomura:2023kwz,Kim:2023jto,Devi:2023vpe,Dasgupta:2021ggp,Nomura:2019lnr,CentellesChulia:2023zhu}, $S_4$~\cite{Penedo:2018nmg,Novichkov:2018ovf,Kobayashi:2019xvz,Liu:2020akv,deMedeirosVarzielas:2023crv,Ding:2021zbg,King:2019vhv}, and $A_5$~\cite{Novichkov:2018nkm,Ding:2019zxk}, along with more expansive groups~\cite{baur2019unification}, double covering of $A_4$ \cite{Mishra:2023cjc,Liu:2019khw,Ding:2023ynd}, $S_4$ \cite{Abe:2023ilq,Abe:2023qmr}, and $A_5$ \cite{Wang:2020lxk,Behera:2022wco,Behera:2021eut}. However, $S_3$, the smallest finite modular group, has not been extensively studied. This will be the focus of our paper.

In this work, we apply modular $S_3$ symmetry to the supersymmetric type-I seesaw model. This scenario, albeit its simplicity, has so far been overlooked in the literature. Since the irreducible representations of $S_3$ are 1- and 2-dimensional, there are many possibilities in assigning lepton chiral supermultiplets into $S_3$ representations. We characterize each different realization of $S_3$ symmetry by the number of 2-dimensional representations employed. We have identified 4 different scenarios, referred to as model A, B, C and D, which are compatible with the current neutrino oscillation data and contain the least number of free parameters.

The manuscript is organized as follows: We briefly describe $S_3$ modular symmetry in Sec.~\ref{sec:S3_symmetry}. In Sec.~\ref{model}, we classify different realizations of $S_3$ modular symmetry which give rise to viable neutrino oscillations. We then explore the neutrino phenomenology for each class of model in Sec.~\ref{sec:Numerics}. Finally, we conclude in Sec.~\ref{sec:conc}.

\section{The $\Gamma_2 \cong S_3$ modular group}
\label{sec:S3_symmetry}
The modular group $\bar\Gamma$ consists of linear fractional transformations, denoted by $\gamma$, acting on the modulus $\tau$ in the upper-half complex plane. The transformation is defined as follows:
\begin{equation}\label{eq:tau-SL2Z}
\tau \longrightarrow \gamma\tau= \frac{a\tau + b}{c \tau + d}\ ,
\end{equation}
where $a$, $b$, $c$, and $d$ are integers satisfying $ad-bc=1$, and $\text{Im} [\tau]>0$. This transformation is isomorphic to $PSL(2,\mathbb{Z})=SL(2,\mathbb{Z})/\{I,-I\}$, where $I$ is the identity transformation. The modular transformation is generated by two fundamental operations, $S$ and $T$:
\begin{eqnarray}
S:\tau \longrightarrow -\frac{1}{\tau}\ , \qquad\qquad
T:\tau \longrightarrow \tau + 1\ ,
\end{eqnarray}
These operations satisfy the algebraic relations:
\begin{equation}
S^2 =\mathbb{I}\ , \qquad (ST)^3 =\mathbb{I}\ ,
\end{equation}
where $\mathbb{I}$ represents the identity transformation.

We define a series of groups, denoted as $\Gamma(N)$ for $N=1,2,3,\dots$, as follows:
\begin{align}
\Gamma(N)= \left \{ 
\begin{pmatrix}
a & b  \\
c & d  
\end{pmatrix} \in SL(2,\mathbb{Z})~ \bigg| ~
\begin{pmatrix}
a & b  \\
c & d  
\end{pmatrix} \equiv
\begin{pmatrix}
1 & 0  \\
0 & 1  
\end{pmatrix} ~~({\rm mod}~ N) \right \}.
\end{align}
For $N=2$, we denote $\bar\Gamma(2)$ as $\Gamma(2)/\{I,-I\}$. Since the element $-I$ is not in $\Gamma(N)$ for $N>2$, we have $\bar\Gamma(N) = \Gamma(N)$, and these are infinite normal subgroups of $\bar\Gamma$, referred to as principal congruence subgroups.

The quotient groups, defined as $\Gamma_N \equiv \bar\Gamma/\bar\Gamma(N)$, are finite modular groups. In these groups, the condition $T^N=\mathbb{I}$ is imposed. Specifically, the groups $\Gamma_N$ with $N=2,3,4,5$ are isomorphic to $S_3$, $A_4$, $S_4$, and $A_5$, respectively \cite{deAdelhartToorop:2011re}.

Modular forms of level $N$ are holomorphic functions $f(\tau)$, which transform under the action of $\Gamma(N)$ as:
\begin{equation}
f(\gamma\tau)= (c\tau+d)^{k_I} f(\tau)~, ~~ \gamma \in \Gamma(N)~ ,
\end{equation}
where $k_I\ge0$ is referred to as the modular weight. For $k_I=0$, the modular form is a constant. Since $T^N=\mathbb{I}$, it follows that $f(\tau+N) = f(\tau)$ and $f(\tau)$ admits a Fourier expansion ($q$-expansion)
\begin{equation}
    f(\tau) = \sum_i a_iq_N^i,\quad\text{where}\quad q_N=e^{2i\pi\tau/N}.
\end{equation}
In this work, we will focus on the group $S_3$, which corresponds to $N=2$.
 
The group $S_3$ exhibits three irreducible representations: the doublet $\mathbf{2}$, the singlet $\mathbf{1}$, and the pseudo-singlet $\mathbf{1'}$. The lowest even-weighted modular forms arise when $k_I=2$ and the corresponding modular coupling is represented as $Y_2^{(\textbf{2})}=(Y_1(\tau),Y_2(\tau))$ and is given in terms of Dedekind eta function ($\eta(\tau)$) as given in Ref.~\cite{Kobayashi:2018vbk}. However, for the practical numerical calculations purpose, we use the $q$-expansion~\cite{Kobayashi:2018vbk} form as given below
\begin{eqnarray} 
Y_1(\tau) &=& \frac 18 + 3q^2 + 3q^4 + 12 q^6 + 3q^8 \cdots ,\nonumber \\
Y_2(\tau) &=& \sqrt 3 q (1+ 4 q^2 + 6 q^{4} + 8 q^{6} \cdots  ),  \label{eq:modular func_S3_expand}
\end{eqnarray}
where $q \equiv e^{i\pi\tau}$.

One crucial thing to note is that the higher-order modular forms  utilized in model building discussed in Sec.~\ref{model} are obtained by applying $S_3$ product rules. Consider two $S_3$ doublets $x=(x_1, x_2)$ and $y=(y_1, y_2)$. The $S_3$ product rule yields $x\otimes y = (x_2y_2 - x_1y_1,x_1y_2+x_2y_1)_2$ $\oplus$ $(x_1y_1+x_2y_2)_1$ $\oplus$ $(x_1y_2-x_2y_1)_{1'}$ where the subscript denote the $S_3$ representation~\cite{Ishimori:2010au}.  Hence, the higher weight modular forms ($Y_\mathrm{a}^{(k_I)}$), where $``\mathrm{a}"$ represents the $S_3$ charge and $k_I$  being the modular weight, used in this paper are shown below
\begin{align}
&Y^{(4)}_{\bf1}=Y^2_1+Y^2_2,\quad
Y^{(6)}_{\bf1}=3Y^2_1Y_2-Y^3_1,\quad
Y^{(6)}_{\bf1'}=3Y_1Y_2^2 - Y_2^3,  \nn\\
&Y^{(4)}_{\bf2}=
\left[\begin{array}{c}
Y^2_2-Y_1^2  \\ 
2Y_1Y_2 \\ 
\end{array}\right] = \left[\begin{array}{c} (Y^{(4)}_{\bf2})_1 \\ (Y^{(4)}_{\bf2})_2 \\ \end{array} \right],\quad
Y^{(6)}_{\bf2}=
\left[\begin{array}{c}
Y^3_1+Y_1Y_2^2 \\ 
Y^3_2+Y_1^2 Y_2 \\ 
\end{array}\right] = \left[ \begin{array}{c}
    (Y^{(6)}_{\bf2})_1  \\
     (Y^{(6)}_{\bf2})_2  \\
\end{array} \right]\nn \\&
Y^{(8)}_{\bf1}=(Y^2_1+Y^2_2)^2.  
\label{eqn:yukawa_c}
\end{align}

\section{Model Framework}
\label{model}
\vspace{0.1cm}

In this section, we describe the models considered in our analysis. Our models are based on the supersymmetric type-I seesaw model in which the minimal supersymmetric Standard Model is extended by 3 electroweak singlet chiral supermultiplets $N^c_i$. To establish our convention and notation, the supermultiplets in the leptonic sector are $L_i(2,-1/2)$, $E^c_i(1,-1)$ and $N^c_i(1,0)$, where the numbers in parenthesis are the electroweak charges, and the subscript $i$ is the flavor index. For completeness, the Higgs supermultiplets are $H_u(1/2,-1/2)$ and $H_d(1/2,1/2)$ with the vev $v_u$ and $v_d$ respectively. The electroweak VEV is $v=\sqrt{v_u^2+v_d^2} = 246$ GeV. For later convenience, we define $\tan\beta = v_d/v_u$.

The superpotential for the lepton sector can be written as \begin{align}
    \mathcal{W} \supset (Y_\ell)_{ij}E^c_iL_jH_d +(Y_\nu)_{ij}N^c_iL_jH_u + \frac{1}{2}(M_R)_{ij}N^c_iN_j^c,
\end{align}
where $Y_\ell$ and $Y_\nu$ are the Yukawa coupling matrices, and $M_R$ is the Majorana mass matrix for $N^c_i$. After electroweak symmetry breaking, the superpotential induces charged lepton masses $M_\ell = Y_\ell v_d/\sqrt{2}$.
The charged lepton mass matrix is diagonalized by unitary matrix $U_\ell$, with $U_\ell^\dagger M_\ell^\dagger M_\ell U_\ell = $ diag($m_e^2,m_\mu^2,m_\tau^2$).
In the neutral lepton sector, $\mathcal{N} = (\nu,N^c)$, the induced mass matrix is given by 
\begin{equation}
    M_\mathcal{N} = \begin{pmatrix}
        0 &M_D^T\\ M_D &M_R
    \end{pmatrix}, 
\end{equation}
where
\begin{equation}
    M_D = \frac{Y_\nu v_u}{\sqrt{2}}.
\end{equation}
In the limit where $|M_R|\gg|M_D|$, the light neutrino mass matrix is given by
\begin{equation}
    M_\nu = M_D^TM_R^{-1}M_D. \label{numass}
\end{equation}
$M_\nu$ can be diagonalized by a unitary matrix $U_\nu$, with $U_\nu^\dagger M_\nu^\dagger M_\nu U_\nu = $ diag.($m_{\nu_1}^2,m_{\nu_2}^2,m_{\nu_3}^2$). The mismatch between $U_\ell$ and $U_\nu$, 
\begin{equation}
    U \equiv U^\dagger_\ell U_\nu, \label{PMNS}
\end{equation}
is the Pontecovo-Maki-Nakagawa-Sakata (PMNS) matrix \cite{Hochmuth:2007wq}.
The modular $S_3$ symmetry is imposed on the leptonic sector. As a result, the structures of $Y_{\ell,\nu}$ and $M_R$ are constrained by modular $S_3$ invariant. Since the $S_3$ symmetry can be realized in many different ways, we systematically classify them by specifying which supermultiplet transforms in the doublet representation of $S_3$. Then, for each class of model, we identify the $S_3$ representation and modular weight of each supermultiplet, which leads to viable neutrino masses and mixing with the least number of free parameters. For our analysis, we focus on scenario where two or more lepton families contain the doublet representation. Having only one lepton species transforming in the doublet generally allows more free parameter, reducing the predictive power of the model. All different modular $S_3$ realizations considered in this work are shown in Table.~\ref{tab:all_models}

\begin{table}[htbp]
\begin{tabular}{||c||c||c|c|c|c|c|c|c|c|c||c||c|c||}
      \hline \hline
     & \textbf{Fields} & $ E_1^c$ & $E_2^c$ & $E_3^c$  &$ L_1$ &$ L_2$ & $L_3$ & $ N_1^c$ &$ N_2^c$ & $N_3^c$ & Free Parameters &\textbf{NO} & \textbf{IO} \\ \hline \hline
       
\multirow{2}{*}{\textbf{MODEL A}} & $S_3$                 & \multicolumn{2}{c|}{2}   &  1 & \multicolumn{2}{c|}{2}  &  1  &\multicolumn{2}{c|}{2}  &  $1$  &\multirow{2}{*}{\large 7}  &\multirow{2}{*}{\textcolor{blue}{\checkmark}} & \multirow{2}{*}{\textcolor{blue}{\checkmark}} \\ \cline{2-11}
                                       &  $k_I$    &  \multicolumn{2}{c|}{1}   & -1 & \multicolumn{2}{c|}{1} & 1 & \multicolumn{2}{c|}{1} & 3  &  &  &  \\ \hline \hline
\multirow{2}{*}{\textbf{MODEL B}} & $S_3$                 & \multicolumn{2}{c|}{2}  & 1 & \multicolumn{2}{c|}{2}  & 1  & 1  & $1$ & $1'$  &\multirow{2}{*}{\large 7}  &\multirow{2}{*}{\textcolor{blue}{\checkmark}} & \multirow{2}{*}{\textcolor{blue}{\checkmark}}  \\ \cline{2-11}
                                       & $k_I$  & \multicolumn{2}{c|}{0}   & 0 & \multicolumn{2}{c|}{2} & 0 & 0 &2 & 0  &  &  &  \\ \hline \hline
\multirow{2}{*}{\textbf{MODEL C}} & $S_3$                 & $1$   & $1'$   & 1 &\multicolumn{2}{c|}{2} & 1  &\multicolumn{2}{c|}{2}  & $1$  &\multirow{2}{*}{\large 4}  &\multirow{2}{*}{\textcolor{red}{\xmark}} & \multirow{2}{*}{\textcolor{blue}{\checkmark}}  \\ \cline{2-11}
                                       & $k_I$                 & 1   & 1   & -1 & \multicolumn{2}{c|}{1} & 1 & \multicolumn{2}{c|}{1} & 1 &   &  & \\ \hline \hline
\multirow{2}{*}{\textbf{MODEL D}} & $S_3$                 & \multicolumn{2}{c|}{2}   & $1'$ &1 & $1'$ & $1'$  &\multicolumn{2}{c|}{2}  & $1'$  &\multirow{2}{*}{\large 9}  &\multirow{2}{*}{\textcolor{blue}{\checkmark}} & \multirow{2}{*}{\textcolor{blue}{\checkmark}}  \\ \cline{2-11}
                                       & $k_I$                 & \multicolumn{2}{c|}{0}   & 0 & 2 & 2 & 0 & \multicolumn{2}{c|}{0} & 4 &   &  & \\ \hline \hline
    \end{tabular}%
  \caption{In this table, we depict the particle content of the different models and their charges under $S_3$ modular symmetry, where, $k_I$ is the modular weight. Also, the number of free real parameters, in addition of the complex modulus $\tau$, and the possible ordering of neutrino masses are provided for each model.
  } 
  \label{tab:all_models}
\end{table}

\subsection{MODEL A}
\label{subsec:model A}
We first consider a scenario where all three lepton species, $E^c_i$, $L_i$, and $N^c_i$, transform in the doublet representation of $S_3$. For definiteness, we take $\Psi \equiv (\Psi_1,\Psi_2)$, where $\Psi = E^c$, $L$ and $N^c$ respectively, as the doublet. The superpotential in the charged lepton sector consistent with modular $S_3$ symmetry is given by
\begin{equation}
	\mathcal{W}_\ell = \alpha_\ell (E^cL)_2Y^{(2)}_2H_d + \beta_\ell (E^cY^{(2)}_2)_1L_3 H_d + \gamma_\ell E^c_3L_3H_d,
 \label{eqn:CL_2}
\end{equation}
where a subscript on the parenthesis indicates its $S_3$ representation.
This superpotential leads to a non-diagonal charged lepton mass matrix
\begin{equation}
	M_\ell = \frac{v_d}{\sqrt{2}}\begin{pmatrix} -{\alpha}_\ell Y_1 &{\alpha}_\ell Y_2 &0\\ {\alpha}_\ell Y_2 &{\alpha}_\ell Y_1 &0\\ {\beta}_\ell Y_1 &{\beta}_\ell Y_2 & \gamma_\ell \end{pmatrix}.
\end{equation}
The Yukawa couplings $\alpha_\ell$, $\beta_\ell$ and $\gamma_\ell$ are in general complex. However, one can perform arbitrary phase redefinition on $E^c_i$ to make them real. Hence, they are completely determined by the charged lepton masses through the relations
 \begin{eqnarray}
	{\rm Tr} \left( M^{}_\ell M^{\dag}_\ell \right) &=& m^{2}_{e} + m^{2}_{\mu} + m^{2}_{\tau} \; ,  \label{eq:tr}\\
	{\rm Det}\left( M^{}_\ell M^{\dag}_\ell \right) &=& m^{2}_{e} m^{2}_{\mu} m^{2}_{\tau} \; , \label{eq:det}\\
	\dfrac{1}{2}\left[{\rm Tr} \left(M^{}_\ell M^{\dag}_\ell\right)\right]^2_{} - \dfrac{1}{2}{\rm Tr}\left[ (M^{}_\ell M^{\dag}_\ell)^2_{}\right] &= & m^{2}_{e}m^{2}_{\mu}+m^{2}_{\mu}m^{2}_{\tau}+m^{2}_{\tau}m^{2}_{e} \; . \label{eq:tr2}
 \label{eqn:charged_identities}
	\end{eqnarray}

For the neutral leptons, the superpotential is given by

\begin{equation}
\begin{aligned}
	\mathcal{W}_\nu &= \alpha_{D} (N^cL)_2Y^{(2)}_2 H_u + \beta_{D} N^c_3(LY^{(4)}_2)_{1}H_u +\gamma_{D} (N^cY^{(2)}_2)L_3H_u + \omega_D N^c_3L_3Y^{(4)}_1H_u\nonumber\\
	&\quad + \alpha_R M (N^cN^c)_2Y^{(2)}_2 + \beta_{R} M N^c_3(N^cY^{(4)}_2)_{1} + M N^c_3N^c_3Y^{(6)}_1.
\label{eqn:neutral_SP_C2}
\end{aligned}
\end{equation}
The first three terms give rise to a Dirac matrix
\begin{align}
	M_D &= \frac{\omega_{D}v_u}{\sqrt{2}}\begin{pmatrix}-\bar{\alpha}_{D} Y_1~&~ \bar{\alpha}_{D} Y_2 ~&~\bar{\gamma}_DY_1\\ 
		\bar{\alpha}_{D} Y_2 ~&~\bar{\alpha}_{D} Y_1~&~ \bar{\gamma}_DY_2\\ 
		 \bar{\beta}_{D}(Y_2^{\textbf{(4)}})_1 ~&~ \bar{\beta}_{D} (Y_2^{\textbf{(4)}})_2  ~&~ Y^{(4)}_1\end{pmatrix},
   \label{eqn:Dirac_MM_C2}
\end{align}
where we introduce a shorthand notation $\bar x = x / \omega_D$. The last three terms of $\mathcal{W}_\nu$ give rise to the Majorana mass matrix for $N^c$,
\begin{align}
	M_R &= M\begin{pmatrix}-2\alpha_{R} Y_1 ~&~ 2\alpha_{R} Y_2 ~&~ \beta_{R} (Y_2^{\textbf{(4)}})_1\\
	2\alpha_{R} Y_2  ~&~ 2\alpha_{R} Y_1 ~&~ \beta_{R}(Y_2^{\textbf{(4)}})_2\\
	\beta_{R}(Y_2^{\textbf{(4)}})_1 ~&~ \beta_{R}(Y_2^{\textbf{(4)}})_2 ~&~ Y_1^{\textbf{(6)}}\end{pmatrix}.
 \label{eqn:Maj_MM_C2}
\end{align}

The above Dirac and Majorana mass matrices lead to the light neutrino mass matrix $M_\nu = M_D^TM_R^{-1}M_D^{}$. Note that the complex combination $m = \omega_D^2v_u^2/M$ sets the scale of light neutrino mass. The phase of $m$ has no effect on neutrino oscillations. Moreover, four of the five parameters $\bar{\alpha}_{D}$, $\bar{\beta}_{D}$, $\bar{\gamma}_D$, $\alpha_R$ and $\beta_R$ can be made real by redefining the phase of $L_i$ and $E^c_i$. Hence, in this scenario, in addition to the modulus $\tau$, the light neutrino oscillations depend on 7 real parameters: $|m|$, $|\bar{\alpha}_{D}|$, $|\bar{\beta}_{D}|$, $|\bar{\gamma}_D|$, $|\alpha_R|$, $|\beta_R|$ and one phase. 

\subsection{MODEL B}
\label{subsec:model B}
In this scenario, we take $E^c = (E^c_1,E^c_2)$ $L = (L_1,L_2)$ to be doublets of $S_3$. The complete $S_3$ charged assignment and the modular weight of each lepton supermultiplet are shown in Model B of Tab.~\ref{tab:all_models}. The superpotential for the charged leptons is given by
\begin{equation}
	\mathcal{W}_\ell = \alpha_\ell (E^cL)_2Y^{(2)}_2H_d + \beta_\ell E_3^c(Y^{(2)}_2 L)_1 H_d + \gamma_\ell E^c_3L_3H_d.
\end{equation}
This leads to a  charged lepton mass matrix
\begin{equation}
	M_\ell = \frac{v_d}{\sqrt{2}}\begin{pmatrix} -{\alpha}_\ell Y_1 &{\alpha}_\ell Y_2 ~&~ 0\\ {\alpha}_\ell Y_2 ~&~{\alpha}_\ell Y_1 &0\\ {\beta}_\ell Y_1 &{\beta}_\ell Y_2 &\gamma_\ell\end{pmatrix}.
\end{equation}
The Yukawa couplings $\alpha_\ell$, $\beta_\ell$, and $\gamma_\ell$ can be made real by an arbitrary phase rotation on the superfields $E^c_i$ and $L_3$. Hence, they are completely determined by charged lepton masses, see equation~\eqref{eqn:charged_identities}.

For the neutral leptons, the superpotential is given by
\begin{align}
	\mathcal{W}_\nu &= \alpha_{D} N_1^c(LY^{(2)}_2)_{1} H_u + \beta_{D} N^c_2(LY^{(4)}_2)_{1}H_u +\gamma_{D} N^c_3(LY^{(2)}_2)_{1'}H_u +\omega_{D} N_1^cL_3H_u\nonumber\\
	&\quad + \alpha_R M N_1^cN_1^c + \beta_{R} MN^c_2 N^c_2 Y^{(4)}_1 +  MN^c_3N^c_3.
 \label{eqn:neutral_SP_C3}
\end{align}
The first four terms in $\mathcal{W}_\nu$ lead to the Dirac mass matrix 
\begin{align}
	M_D &= \frac{\omega_Dv_u}{\sqrt{2}}\begin{pmatrix}\bar{\alpha}_{D} Y_1 ~&~ \bar{\alpha}_{D} Y_2 ~&~1\\ 
		\bar{\beta}_{D} (Y_2^{\bf(4)})_1 ~&~\bar{\beta}_{D} (Y_2^{\bf(4)})_2~&~ 0\\ 
           \bar{\gamma}_{D}Y_2 ~&~-\bar{\gamma}_{D}Y_1  ~&~0\end{pmatrix},
   \label{eqn:Dirac_MM_C3}
\end{align}
where we have employed a short-hand notation $\bar x = x/\omega_D$. The last three terms of $\mathcal{W}_\nu$ are the Majorana masses for $N^c_i$
\begin{align}
	M_R &= M \begin{pmatrix}
	    \alpha_{R} ~&~ 0 ~&~ 0\\
        0 ~&~ \beta_{R} Y_1^{\bf(4)} ~&~ 0\\
        0 ~&~ 0 ~&~ 1
	\end{pmatrix}.
 \label{eqn:Maj_MM_C3}
\end{align}

As is the case for Model A, the complex combination $m=\omega_D^2v_u^2/M$ determines the scale of light neutrino masses. Also, four of the five complex parameters, $\bar{\alpha}_D$, $\bar{\beta}_D$, $\bar{\gamma}_D$, $\alpha_R$ and $\beta_R$, can be made real by a phase redefinition of $L_i$ and $N^c_i$. Hence, in addition to the modulus $\tau$, there are 7 free real parameters in the light neutrino sector: $|m|$, $|\bar{\alpha}_D|$, $|\bar{\beta}_D|$, $|\bar{\gamma}_D|$, $|\alpha_R|$, $|\beta_R|$ and one phase.

\subsection{MODEL C}
\label{subsec:MODEL C}
In this scenario, we take $L=(L_1,L_2)$ and $N^c = (N^c_1,N^c_2)$ as $S_3$ doublets. The complete $S_3$ charge assignment and modular weight can be found in Model C of Tab.~\ref{tab:all_models}.

The superpotential for the charged lepton sector is
\begin{equation}
    \mathcal{W}_\ell = \alpha_\ell E^c_1 (L Y_2^{(2)})_{1}  H_d + \beta_\ell E^c_2 (L_2Y_2^{(2)})_{1'} H_d + \gamma_{\ell} E_3^c L_3 H_d
\end{equation}
This leads to a charged lepton mass matrix
\begin{equation}
    M_\ell = \frac{v_d}{\sqrt{2}}\begin{pmatrix}
        \alpha_\ell Y_1 & \alpha_\ell Y_2 & 0\\
        \beta_\ell Y_2 & -\beta_\ell Y_1 &0\\
        0 &0 &\gamma_\ell
    \end{pmatrix}.
\end{equation}
Again, as in the case of Model A and Model B above, the couplings $\alpha_\ell$, $\beta_\ell$ and $\gamma_\ell$ can be made real by redefining the phase of $E^c_i$. Hence they are completely determined by the charged lepton masses via equation~\eqref{eqn:charged_identities}.

In the neutral lepton sector, the superpotential is given by
\begin{equation}
\begin{aligned}
    \mathcal{W}_\nu &= \alpha_D \left[(N^c L)_2 Y_2^{(2)}\right]_{1}  H_u + \beta_D  (N^c Y_2^{(2)})_{1} L_3 H_u + \omega_{D} N_3^c (L Y_2^{(2)})_1 H_u \\
    &\quad+ M\left(\left[(N^c N^c)_2Y^{(2)}_2\right]_1 + \alpha_R N_3^c (N^c Y_2^{(2)})_1\right).
\end{aligned}
\end{equation}
It leads to the Dirac mass matrix
\begin{equation}
    M_D = \frac{\omega_Dv_u}{\sqrt{2}}\begin{pmatrix}
        -\bar{\alpha}_D Y_1 & \bar{\alpha}_D Y_2 & \bar{\beta}_D Y_1\\
        \bar{\alpha}_D Y_2 & \bar{\alpha}_D Y_1 & \bar{\beta}_D Y_2\\
        Y_1 & Y_2 &0
    \end{pmatrix},\label{eqn:Dirac_MM_C5b}
\end{equation}
and the Majorana mass matrix
\begin{equation}
    M_R = M\begin{pmatrix}
        - Y_1 & Y_2 &  \alpha_R Y_1\\
        Y_2 & Y_1 &  \alpha_R Y_2\\
        \alpha_R Y_1 & \alpha_R Y_2 &0
    \end{pmatrix}.\label{eqn:Maj_MM_C5b}
\end{equation}
Light neutrino mass scale in Model C, similar to Model A and B, depends on the combination $m=\omega_D^2v_u^2/M$. Moreover, by performing a phase redefinition of $L_i$ and $N^c_i$, one can make all the complex parameters $\bar{\alpha}_D$, $\bar{\beta}_D$ and $\alpha_R$ real. Hence, in addition to the complex modulus $\tau$, neutrino oscillations are determined by 4 real parameters:  $|m|$, $|\bar{\alpha}_D|$, $|\bar{\beta}_D|$ and $|\alpha_R|$.

\subsection{Model D}
\label{subsec:model D}
In this scenario, we take $E^c=(E^c_1,E^c_2)$ and $N^c = (N^c_1,N^c_2)$ as $S_3$ doublets. The complete $S_3$ charge assignment and modular weight can be found in Model D of Tab.~\ref{tab:all_models}.

The superpotential in the charged lepton sector can be written as
\begin{equation}
    \mathcal{W}_\ell = \alpha_\ell (E^c Y_2^{(2)})_{1} L_1 H_d +  \beta_\ell (E^c Y_2^{(2)})_{1'} L_2 H_d + \gamma_\ell E_3^c  L_3 H_d 
\end{equation}
The above superpotential
leads to a charged lepton mass matrix
\begin{equation}
    M_\ell =  \frac{v_d}{\sqrt{2}}\begin{pmatrix}
        \alpha_\ell Y_1 & \beta_\ell Y_2 & 0\\
        \alpha_\ell Y_2 & -\beta_\ell Y_1 & 0\\
        0 & 0 & \gamma_\ell
    \end{pmatrix}.
\end{equation}
As in the previous scenarios, the couplings $\alpha_\ell$, $\beta_\ell$ and $\gamma_\ell$ can be made real by arbitrary phase redefinition of the $E^c_i$ and one of the $L_i$ supermultiplet. Thus, they are determined by the charged lepton masses via equation~\eqref{eqn:charged_identities}.

In the neutral lepton sector, the superpotential is given by
\begin{equation}
\begin{aligned}
    \mathcal{W}_\nu &= \omega_D  (N^cY^{(2)}_2)_{1} L_1 H_u + \alpha_D  (N^cY^{(2)}_2)_{1'} L_2 H_u + \beta_{D} N_3^c L_1 Y_{1'}^{(6)} H_u + \gamma_{D} N_3^c L_2 Y_{1}^{(6)} H_u \\
    &\quad+ \eta_{D} N_3^c L_3 Y_{1}^{(4)} H_u 
    + M \left[ (N^cN^c)_1 + \alpha_{R} N^c_3(N^cY^{(4)}_2)_{1'} + \beta_R N^c_3N^c_3Y^{(8)}_1\right].
\end{aligned}
\end{equation}
The first 5 terms in the superpotential give rise to a Dirac mass matrix
\begin{equation}
    M_D = \frac{\omega_Dv_u}{\sqrt{2}}\begin{pmatrix}
        Y_1 & \bar{\alpha}_D Y_2 &0 \\
        Y_2 & -\bar{\alpha}_D Y_1 &0 \\
        \bar{\beta}_{D}Y_{1'}^{(6)} &\bar{\gamma}_{D}Y_1^{(6)} &\bar{\eta}_DY_1^{(4)}
    \end{pmatrix},
\end{equation}
and the last 3 terms give the Majorana mass matrix
\begin{equation}
    M_R = M\begin{pmatrix}
        1 & 0& \alpha_{R}(Y^{(4)}_2)_2\\
        0 &1 & -\alpha_{R}(Y^{(4)}_2)_1\\
        \alpha_{R}(Y^{(4)}_2)_2 &-\alpha_{R}(Y^{(4)}_2)_1 &\beta_RY^{(8)}_1
    \end{pmatrix}.
\end{equation}

Concerning the light neutrinos, the combination $m=\omega_D^2v_u^2/M$ set their masses scale. Moreover, one can redefine the phase of $L_i$ and $N^c_i$ to rotate away all but two phases in the complex parameters $\bar{\alpha}_D$, $\bar{\beta}_D$ $\bar{\gamma}_{D}$, $\bar{\eta}_{D}$, $\alpha_R$ and $\beta_R$. Hence, in this model the light neutrino phenomenology depends on 9 real parameters: $|m|$, $|\bar{\alpha}_D|$, $|\bar{\beta}_D|$ $|\bar{\gamma}_{D}|$, $|\bar{\eta}_{D}|$, $|\alpha_R|$, $|\beta_R|$ and two phases. The large number of free parameters in this model makes it somewhat less interesting than the other models considered above. However, we find that in this scenario, neutrino masses are compatible with both normal ordering and inverted ordering.
\section{Neutrino Phenomenology}
\label{sec:Numerics} 
In this section, we numerically analyze the neutrino mass matrices depicted in the previous section. This is achieved by performing a scan on the model parameter space. For each model, we identify the region of parameter space consistent with neutrino oscillation data at the 2-$\sigma$ level, see Table~\ref{table:nufit}. In our analysis, the solar mass splitting is taken to be $\Delta m^2_{sol} = m_{\nu_2}^2-m_{\nu_1}^2$, while the atmospheric mass splitting is $\Delta m^2_{atm} = m_{\nu_3}^2-m_{\nu_1}^2$ ($m_{\nu_2}^2-m_{\nu_3}^2$) for NO (IO).

\begin{table}[htbp]
\centering
\renewcommand{\arraystretch}{1.2}
\begin{tabular}{|lcc|} 
\hline
Parameter$\qquad\qquad$ & \multicolumn{2}{c|}{Best-fit value and 1-$\sigma$ range} \\ 
\hline \hline

%
& \textbf{NO} & \textbf{IO} \\
%
$\Delta m^2_\text{sol}/(10^{-5}\text{ eV}^2)$ & $7.41^{+0.21}_{-0.20}$ & $7.41^{+0.21}_{-0.20}$ \\
%
$\Delta m^2_\text{atm}/(10^{-3}\text{ eV}^2)$ & $2.507^{+0.028}_{-0.027}$ & $2.486^{+0.025}_{-0.028}$ \\
%
%
$\sin^2\theta_{12}$ & $0.303^{+0.012}_{-0.011}$ & $0.303^{+0.012}_{-0.012}$ \\
%
$\sin^2\theta_{13}$ & $0.02225^{+0.00056}_{-0.00059}$ & $0.02223^{+0.00058}_{-0.00058}$ \\
%
$\sin^2\theta_{23}$ & $0.451^{+0.019}_{-0.016}$ & $0.569^{+0.016}_{-0.021}$ \\
%
$J_\text{CP}$ & $-0.0254^{+0.0115}_{-0.0080}$  & $-0.0330^{+0.0044}_{-0.0011}$ \\ \hline \hline
%
\end{tabular}
\caption{The best-fit values for the neutrino oscillation parameters and their 1-$\sigma$ ranges as extracted from Ref.\cite{Esteban:2020cvm}.  
}
\label{table:nufit}
\end{table}

In our scan, the parameters $\bar{\alpha}_D$, $\bar{\beta}_D$, $\bar{\gamma}_D$, $\bar{\eta}_D$, $\alpha_R$ and $\beta_R$ are varied within the range [$10^{-4},10^4$]. The combination $\omega_D^2v_u^2/M$, which sets the neutrino mass scale, is taken to be in the range (0, 1 eV]. The modulus $\tau$ is varied within the fundamental domain defined by 
\begin{equation}
    \text{Im}(\tau) >0,\quad \Big|\text{Re}(\tau)\Big|\leq \frac12,\:~\text{and }\: |\tau| \geq 1.
\end{equation}
For each set of parameters, we first numerically diagonalize both the charged lepton and the neutrino mass matrices to obtain the PMNS matrix $U$ and the mass splitting $\Delta m^2_{sol},\Delta m^2_{atm}$. Then, the oscillation angles are extracted from
\begin{equation}
\sin^2 \theta_{13}= |U_{13}|^2,\quad\sin^2 \theta_{12}= \frac{|U_{12}|^2}{1-|U_{13}|^2},\quad\sin^2 \theta_{23}= \frac{|U_{23}|^2}{1-|U_{13}|^2},
\end{equation}
and the Jarlskog invariant $(J_{CP})$ is determined from
\begin{equation}
    J_{CP} = \text{Im}\left(U_{12}^{}U_{23}^{}U_{13}^\ast U_{22}^\ast\right).
\end{equation}

\begin{figure}[htpb]
\centering
\includegraphics[scale=.7]{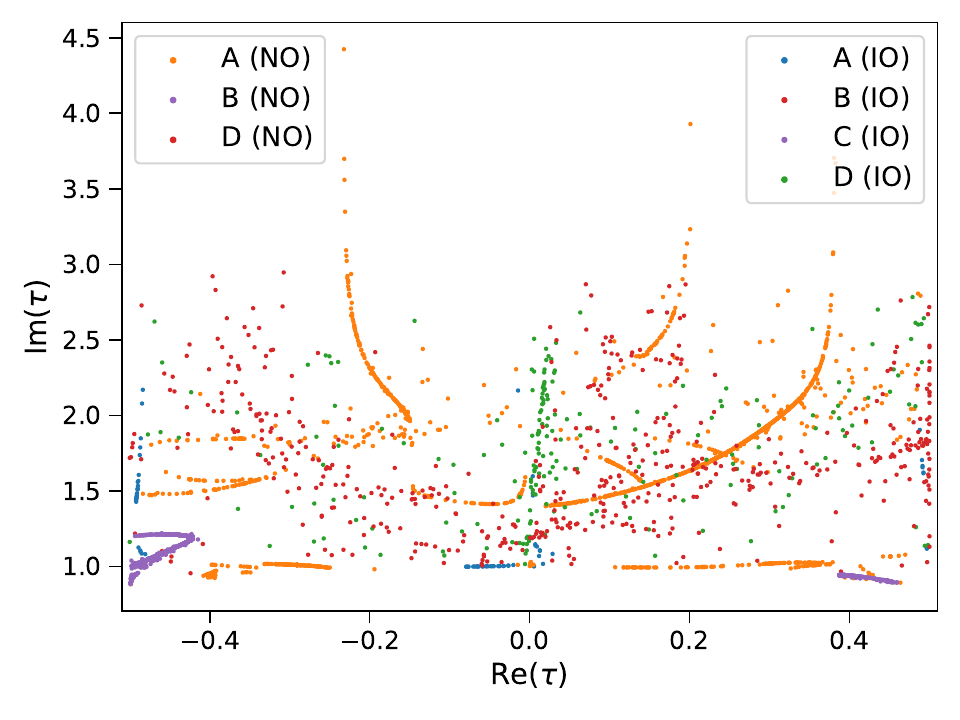}
\caption{The modulus $\tau$ that leads to viable neutrino oscillation parameters for Model A, B, C, and D.}
\label{fig:tau}
\end{figure}

Our analysis shows that Models A, B, and D can accommodate both normal ordering (NO) and inverted ordering (IO) neutrino masses. However, Model C is only compatible with IO due to its minimal structure. Fig.~\ref{fig:tau} shows the modulus $\tau$ in each model which produces neutrino oscillation parameters within 2-$\sigma$ of the best-fit values. It is interesting to note that for both NO and IO in Model B, the modulus tends to clusters in the bottom left or bottom right regions of the fundamental domain. Similarly, for Model C (IO), the modulus is cluster around the bottom left corner of the fundamental domain.

In addition to the oscillation parameters, there are constraints on neutrino masses from cosmology and direct searches. The sum of neutrino masses is constrained by the cosmic microwave background and the baryonic acoustic oscillations measurements, with $\sum_i m_{\nu_i}\lesssim 0.12$-0.52 eV at 95\% confidence level (CL)~\cite{Planck:2018vyg}. The range in the upper limit reflects uncertainties due to the assumed cosmological models. These limits translate to the upper bound on the lightest neutrino mass, $m_{light} \lesssim 0.02$-0.16 eV. For the models considered in this paper, the NO scenarios are compatible with the most stringent upper bound $m_{light}\lesssim0.02$ eV. However, the IO scenarios are only compatible with the more relaxed constraint $m_{light}\lesssim0.16$ eV. The range of $m_{light}$ in each scenario is shown in Tab.~\ref{tab:numassobservables}.

\begin{table}[htbp]
\centering
\begin{tabular}{|c|c|c|c|c|}\hline
& Ordering & $m_{light}$ [eV] &$m^{eff}_{\nu_e}$ [eV] &$m^{eff}_{\beta\beta}$ [eV]\\\hline
\multirow{2}{*}{\textbf{MODEL A}} & NO & $(0.00064,0.12)$ &$(0.0088,0.12)$ &$(0.00033,0.057)$\\
& IO &$(0.058,0.22)$ &$(0.058,0.22)$ &$(0.026,0.11)$\\\hline
\multirow{2}{*}{\textbf{MODEL B}} & NO &$(0.0023,0.077)$ &$(0.0090,0.077)$ &$(0.0016,0.077)$ \\
& IO &$(0.050,0.14)$ &$(0.049,0.13)$ &$(0.048,0.13)$\\\hline
\textbf{MODEL C} & IO &$(0.071,0.072)$ &$(0.071,0.072)$ &$(0.055,0.062)$\\\hline
\multirow{2}{*}{\textbf{MODEL D}} & NO &$(0.0075,0.11)$ &$(0.012,0.11)$ &$(0.0082,0.10)$\\
& IO &$(0.050,0.15)$ &$(0.050,0.15)$ &$(0.022,0.14)$\\\hline
\end{tabular}
\caption{The range of $m_{light}$, $m_{\nu_e}^{eff}$ and $m^{eff}_{\beta\beta}$ in each scenario.}
\label{tab:numassobservables}
\end{table}

\begin{figure}[htpb]
\centering
\includegraphics[scale=.7]{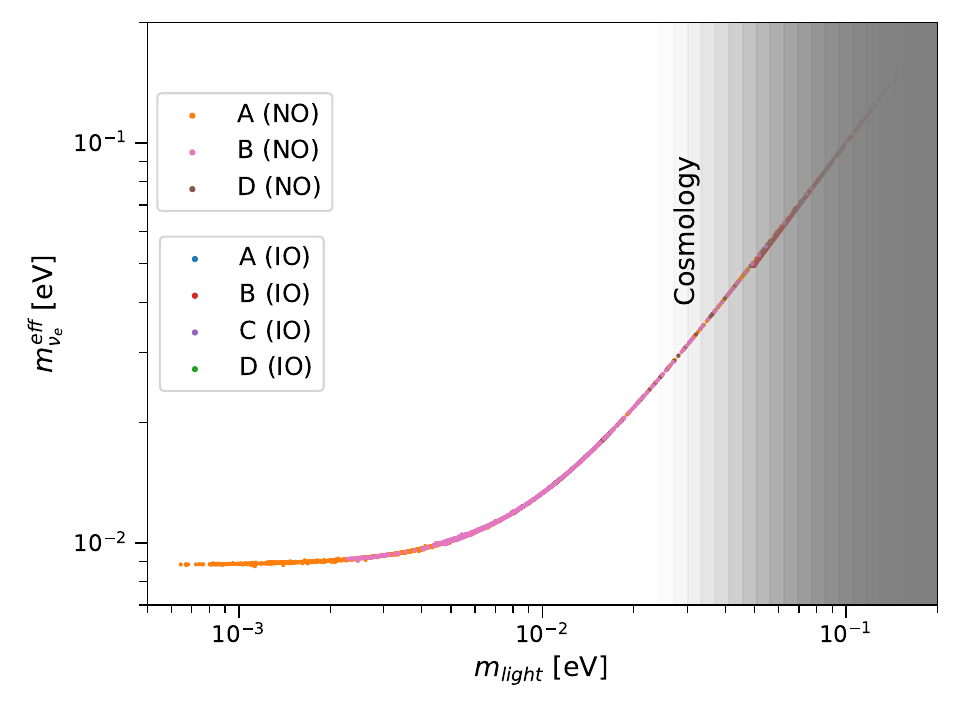}
\caption{The effective electron neutrino mass as a function of the lightest neutrino mass. The vertical shaded region is excluded by cosmological measurements.}
\label{fig:mnuvsmlight}
\end{figure}

\begin{figure}[htpb]
\centering
\includegraphics[scale=.7]{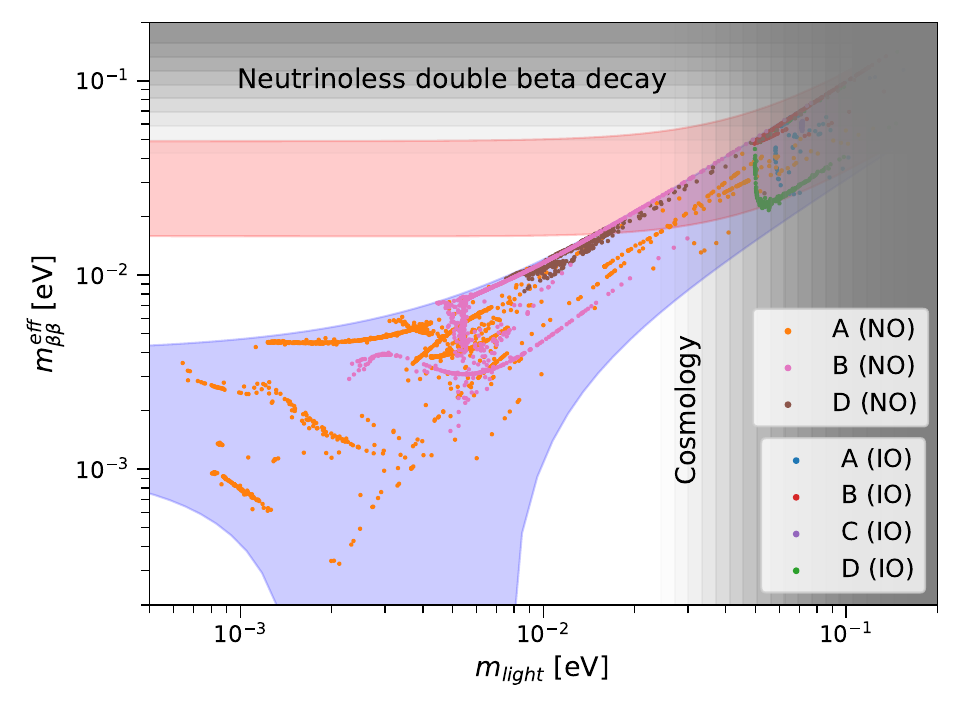}
\caption{The effective Majorana mass for $\nu_e$ as a function of the lightest neutrino mass. The vertical shaded region is excluded by cosmological measurements. The horizontal shaded region is excluded by the neutrino-less double beta decay experiments.}
\label{fig:meevsmlight}
\end{figure}

For direct probes of neutrino mass, the effective electron neutrino mass, defined as $m^{eff}_{\nu_e} = \sqrt{\sum_i |U_{ei}|^2m_{\nu_i}^2}$, is determined from the endpoint of beta decay spectrum. The most stringent upper bound on $m^{eff}_{\nu_e}$ is provided by the KATRIN experiment, with $m^{eff}_{\nu_e}\lesssim0.8$ eV at 90\% CL~\cite{KATRIN:2021uub}. This upper bound is roughly an order of magnitude above the expected $m^{eff}_{\nu_e}$ for Model A, B, C, and D, see Fig.~\ref{fig:mnuvsmlight}. On the other hand, the neutrino-less double beta decay experiments probe the effective Majorana mass of $\nu_e$, defined by $m^{eff}_{\beta\beta}=|\sum_i U_{ei}^2m_{\nu_i}|$. The strongest bound on $m^{eff}_{\beta\beta}$ is provided by the KamLAND-Zen experiment, with $m^{eff}_{\beta\beta}\le0.036$--0.156 eV at 90\% CL~\cite{KamLAND-Zen:2022tow}. The range in the upper bounds reflects uncertainties in nuclear matrix element estimations. With the most stringent estimation of the bound, the IO scenario of Model B and C are ruled out; see Fig~\ref{fig:meevsmlight}. For convenience, possible values of $m^{eff}_{\nu_e}$ and $m^{eff}_{\beta\beta}$ in each scenario are shown in Tab.~\ref{tab:numassobservables}. 

\section{Conclusion and Discussion}
\label{sec:conc}
We have constructed viable models of neutrino masses with the $S_3$ modular symmetry in a type-I super-symmetric framework. Our approach introduces three $SU(2)_L$ singlet chiral supermultiplet ($N^c$) to facilitate the type-I seesaw mechanism. The Yukawa couplings are constructed using the $S_3$ modular forms. Flavor $S_3$ symmetry is broken by the vev of the complex modulus $\tau$. This discrete symmetry proves instrumental in determining the structure of the neutrino mass matrix. We have identified 4 different realizations of the modular $S_3$ symmetry. All four of them are compatible with IO, while Models A, B, and D are also compatible with NO. 

We perform a parameter scan for all 4 models to identify the model parameter space consistent with neutrino oscillation data at the 2-$\sigma$ level. We then determined the lightest neutrino mass, the effective electron neutrino mass, and the effective Majorana mass of $\nu_e$ in each scenario and compared them against experimental bounds. It turns out the IO scenario of each model is tightly constrained by cosmological bound on the lightest neutrino mass. Moreover, the IO scenarios of Model B and C are also strongly constrained by the effective Majorana mass. On the other hand, the effective electron neutrino mass in all scenarios considered in this work is well below the current experimental limit. 

The next generation of neutrino mass measurement can probe the parameter space of the model proposed in this work even further. On the cosmology front, the Simon Observatory can measure the sum of neutrino mass with a sensitivity of 0.04 eV~\cite{SimonsObservatory:2019qwx}. This will cover the entire parameter space of all models considered in this work. For direct measurements, the planned phase-II of the LEGEND experiment is projected to constrain $m_{\beta\beta}^{eff}\le0.013$--0.029 eV~\cite{LEGEND:2017cdu}. This will cover all the parameter space in the IO scenarios and a large portion of parameter space in the NO cases. On the other hand, KATRIN and HOLMES experiments are projected to provide the limit $m^{eff}_{\nu_e}<0.20$ eV~\cite{Aker:2019uuj,Alpert:2014lfa}, which only cover a small portion of the IO scenario of Model A parameter space.

\begin{acknowledgments}
MKB and PI acknowledge support from the NSRF via the Program Management Unit for Human Resources \& Institutional Development, Research, and Innovation [grant no. B13F660066]. The work of PU is supported in part by the Mid-Career Research Grant from the National Research Council of Thailand under contract no. N42A650378. MKB and PU also acknowledge the National Science and Technology Development Agency, National e-Science Infrastructure Consortium, Chulalongkorn University, and the Chulalongkorn Academic Advancement into Its 2nd Century Project (Thailand) for providing computing infrastructure that has contributed to the results reported within this paper. CP is supported by Fundamental Fund 2567 of Khon Kaen University and Research Grant for New Scholar, Office of the Permanent Secretary, Ministry of Higher Education, Science, Research and Innovation under contract no. RGNS 64-043.
\end{acknowledgments}

\textbf{Note added.} While this work is being completed, Ref.~\cite{Marciano:2024quu} appears. The work features a seesaw mechanism under modular $S_3$ symmetry, employing two right-handed neutrinos and predicting a normal ordering of neutrino masses.

\bibliography{s3_type1} 
\bibliographystyle{unsrt}
\end{document}